\documentclass[12pt]{article}
\begin{document}

\def\ba{\begin{eqnarray}}
\def\ea{\end{eqnarray}}

\begin{titlepage}
\title{Possible  effects of space-time
non-metricity on neutrino oscillations}
\author{ M. Adak  \\
 {\small Department of Physics, Pamukkale University,}\\
{\small 20100 Denizli, Turkey} \\ {\small madak@pamukkale.edu.tr} \\ \\
T. Dereli \\
 {\small Department of Physics, Ko\c{c} University,}\\
 {\small 34450 Sar{\i}yer-\.{I}stanbul, Turkey  }  \\
{\small tdereli@ku.edu.tr} \\   \\
L. H. Ryder\\
{\small Department of Physics, University of Kent,}\\
{\small  Canterbury, Kent CT2 7NF, UK}\\
{\small l.h.ryder@ukc.ac.uk}}
\vskip 1cm
\date{   }
\maketitle

\begin{abstract}

 \noindent
The contribution of gravitational neutrino oscillations to the
solar neutrino problem is studied by constructing  the Dirac
Hamiltonian and calculating the corresponding dynamical phase in
the vicinity of the Sun in a non-Riemann background Kerr
space-time with torsion and non-metricity.  We show that certain
components of non-metricity and the axial as well as non-axial
components of torsion may contribute to neutrino oscillations. We
also note that the rotation of the Sun may cause a suppression of
transitions among neutrinos. However, the observed solar neutrino
deficit could not be explained by any of these effects because
they are of the order of Planck scales.

\end{abstract}
\end{titlepage}

\section{Introduction}

Neutrinos always attracted a lot of attention in high energy
physics~\cite{hax}. A major problem of interest at present is the
solar neutrino problem. The Sun is a strong source of electron
neutrinos $\nu_e $ because of the thermonuclear reactions taking
place in its core. According to the standard solar model, the
number of $\nu_e$ to be emitted from the Sun can be predicted. At
the same time, the flux of electron neutrinos coming from the Sun
can be measured on earth. The measured amount of $\nu_e$ is
approximately one third of the predicted amount. Essentially, this
is the so-called {\it solar neutrino problem}. One well known
solution to this problem is provided by the assumption of {\it
neutrino oscillations}~\cite{hax},\cite{bil2}. Briefly stated, the
neutrino oscillations imply that the electron neutrinos coming out
of the Sun may be converted to other neutrino species, muon
$\nu_\mu $ and tau $\nu_\tau $ during their journey towards the
earth, assuming neutrinos to have a mass whereas the standard
electroweak model asserts zero mass for them. It should also be
noted that all the above arguments have been cast in Minkowski
space-time. However, we know that we live in a curved space-time
-- perhaps even in a curved space-time with torsion and
non-metricity. Therefore, in more recent years, physicists have
turned their attention to specifically gravitational contributions
to neutrino oscillations -- see
\cite{ahl},\cite{wudka},\cite{lipkin},\cite{car},\cite{kon} and
references therein. We recently investigated the effects of
space-time torsion on neutrino oscillations~\cite{muz1}- and see
also \cite{gasperini},\cite{alimoh}. The essence of this work is
to calculate the dynamical phase of neutrinos, by finding the form
of the Hamiltonian, H, from the Dirac equation in a non-Riemannian
space-time. The phase then follows from the formula
 \ba
   i\hbar \frac{\partial \psi }{\partial t } = H \psi \;\; \Longrightarrow \;\;
             \psi (t) = e^{-\frac{i}{\hbar }\int{H dt} } \psi (0)
             \label{1}
 \ea
where $\psi $ is a Dirac 4-spinor and $H$ is a $4\times 4$ matrix.
The Hamiltonian $H$ will depend, for example, on momentum
$\vec{p}$, and this is expressed not as a differential operator
but simply as a vector.\footnote{The exponential of $ e^{-\frac{i
}{\hbar}\int{H dt}} $ is defined by its power series expansion.}
In this note we investigate within the same approach the  possible
effects of space-time non-metricity on neutrino oscillations.

\section{Space-Time Geometry}

Space-time is denoted by the triple $ \{M,g,\nabla \} $ where M is
a 4-dimensional differentiable manifold, equipped with a
Lorentzian metric $ g $ which is a (0,2)-type covariant,
symmetric, non-degenerate tensor and $ \nabla $ is a connection
which defines parallel transport of vectors (or more generally
tensors). We shall give a coordinate system set up at a point, $ p
\in M $, by coordinate functions (or independent variables) $ \{
x^\alpha (p) \} $, $ \alpha =\hat{0},\hat{1},\hat{2},\hat{3}$.
This coordinate system forms a set of {\it natural} (or {\it
coordinate}) {\it reference frame} at $ p $ as $ \{
\frac{\partial}{\partial x^\alpha }(p) \} $, with shorthand
notation $ \partial_\alpha \equiv \frac{\partial }{\partial
x^\alpha } $. This natural reference frame is a basis vector set
for the tangent space at $ p $, denoted by $ T_p(M) $. Similarly,
differentials $ \{ dx^\alpha (p) \} $ of coordinate functions $ \{
x^\alpha (p) \} $ at $ p $, form a {\it natural} (or {\it
coordinate}) {\it reference co-frame} in the co-tangent space at $
p $, denoted by $ T^*_p(M) $. Interior product of the basis
vectors with the basis co-vectors is defined by the Kroenecker
symbol:
 \ba
      dx^\alpha (\frac{\partial}{\partial x^\beta })  \equiv
     \imath_{\partial_\beta} dx^\alpha = \delta^\alpha_\beta \; . \label{11}
 \ea
In general, any set of linearly independent  vectors in tangent
space, $ T_p(M) $, can be taken as basis vectors and these vectors
can be orthonormalized by, for example, the Gram-Schmidt process.
We denote a set like this by $ \{ X_a  \} $, $a = 0,1,2,3$ and
call it an {\it orthonormal reference frame}. In this case the
metric defined on M satisfies the relation
 \ba
 g(X_a,X_b) = \eta_{ab}
 \ea
where $ \eta_{ab} $ is known as the Minkowski metric which is a
matrix whose diagonal terms are -1,1,1,1 and off-diagonal terms
are zero. The basis set dual to the orthonormal reference frame
are denoted by $ \{ e^a \} $, a = 0,1,2,3, and called the {\it
orthonormal reference co-frame}. $ \{ X_a \} $ and its dual $ \{
e^a \} $ satisfy the following set of equalities that is another
manifestation of eqn.(\ref{11}):
 \ba
    e^a(X_b) \equiv \imath_{X_b}(e^a) = \delta^a_b \; . \label{12}
 \ea
Here we adhere to the following conventions: indices denoted by
Greek letters $\alpha$, $\beta , \;\; \cdots =
\hat{0},\hat{1},\hat{2},\hat{3}$ and $\mu$, $\nu , \;\; \cdots =
\hat{1},\hat{2},\hat{3}$ are holonomic or coordinate indices, $a$,
$b, \;\; \cdots = 0,1,2,3$ and $i$, $j, \;\; \cdots = 1,2,3$ are
anholonomic or frame indices. In terms of the local coordinate
frame $  \partial_\alpha (p) $, the orthonormal frame $  X_a (p) $
can be expanded via the so-called vierbein (or tetrad) $
{h^\alpha}_a (p) $ as
 \ba
    X_a (p) ={h^\alpha}_a (p) \partial_\alpha (p) \; .
 \ea
In order for $ X_a $ to serve as an anholonomic  basis, the $
{h^\alpha}_a (p) $ are required to be non-degenerate, i.e., $
\mbox{det} {h^\alpha}_a (p) \neq 0 $. In $ T^*_p (M) $ an
orthonormal co-frame $ e^a (p)$ can be expanded in terms of the
local coordinate co-frame $ dx^\alpha (p) $ as
 \ba
    e^b (p)= {h^b}_\beta (p) dx^\beta (p) \; .
 \ea
The inverse vierbein $ {h^b}_\beta (p) $ have to be non-degenerate
as well. Moreover, the duality of the frame and the co-frame
requires for the vierbein and its inverse to satisfy
 \ba
   \imath_{X_a}e^b = {h^\alpha}_a (p) {h^b}_\alpha (p) = \delta^b_a \; .
 \ea
We set the space-time orientation  by the choice $\epsilon_{0123}=
1$. The non-metricity 1-forms, torsion 2-forms and curvature
2-forms are defined by the Cartan structure equations
 \ba
   2Q_{ab} &=& -D\eta_{ab} := \Lambda_{ab} + \Lambda_{ba} \; \; , \label{nonmet}\\
   T^a &=& De^a := de^a + {\Lambda^a}_b \wedge e^b \; \; , \label{torsion}\\
  {R^a}_b &=& D{\Lambda^a}_b  :=
     d{\Lambda^a}_b +{\Lambda^a}_c \wedge {\Lambda^c}_b \;
     .\label{curva}
 \ea
$ d,\quad D, \quad \imath_a,\quad *$ denote the exterior
derivative, the covariant exterior derivative, the interior
derivative and the Hodge star operator, respectively. The linear
connection 1-forms can be decomposed in a unique way according
to~\cite{der1}:
 \ba
  {\Lambda^a}_b = {\omega^a}_b + {K^a}_b + {q^a}_b + {Q^a}_b \;  \label{connec}
 \ea
where $ {\omega^a}_b $ are the Levi-Civita connection 1-forms
 \ba
    {\omega^a}_b \wedge e^b = -de^a \; ,
 \ea
$ {K^a}_b $ are the contortion 1-forms
 \ba
  {K^a}_b \wedge e^b = T^a \label{contor}
 \ea
and $ {q^a}_b $ are the anti-symmetric tensor 1-forms
 \ba
     q_{ab} = -(\imath_a Q_{bc}) e^c
        + (\imath_b Q_{ac}) e^c \; .\label{antisy}
 \ea

It is cumbersome to take into account all components of
non-metricity and  torsion  in gravitational models. Therefore we
will be content with dealing  only with certain irreducible parts
of them to gain physical insight. The irreducible decompositions
of torsion and non-metricity  invariant under the Lorentz group
are summarily given below. For details one may consult
Ref.~\cite{heh}. The non-metricity 1-forms $ Q_{ab} $ can be split
into their trace-free $ \overline{Q}_{ab} $ and the trace parts as
 \ba
   Q_{ab} = \overline{Q}_{ab} + \frac{1}{4} \eta_{ab}Q          \label{118}
 \ea
where the Weyl 1-form $Q={Q^a}_a$ and $ \eta^{ab}\overline{Q}_{ab}
= 0 $. Let us define
 \ba
    \Lambda_b &:=& \imath_a { \overline{Q}^a}_b, \;\;\;\;\;\;\;\;\;\;
                   \Lambda := \Lambda_a e^a,    \nonumber  \\
    \Theta_b &:=& {}^*(\overline{Q}_{ab} \wedge e^a), \;\;\;
    \Theta := e^b \wedge \Theta_b, \;\;\;
    \Omega_a := \Theta_a -\frac{1}{3}\imath_a\Theta    \label{119}
 \ea
as to use them in the decomposition of $ Q_{ab} $ as
 \ba
     Q_{ab} = Q_{ab}^{(1)} + Q_{ab}^{(2)} +
            Q_{ab}^{(3)} +Q_{ab}^{(4)}             \label{120}
 \ea
where
 \ba
    Q_{ab}^{(2)} &=& \frac{1}{3} {}^*(e_a \wedge \Omega_b +e_b
    \wedge \Omega_a)   \\
    Q_{ab}^{(3)} &=& \frac{2}{9}( \Lambda_a e_b +\Lambda_b e_a
    -\frac{1}{2} \eta_{ab} \Lambda )  \\
    Q_{ab}^{(4)} &=& \frac{1}{4} \eta_{ab} Q    \\
    Q_{ab}^{(1)} &=& Q_{ab}-Q_{ab}^{(2)}
                 - Q_{ab}^{(3)} -Q_{ab}^{(4)} \;.
 \ea
We have $ \imath^a Q_{ab}^{(1)} =\imath^a Q_{ab}^{(2)} =0,
\;\;\;\;\;  \eta^{ab}Q_{ab}^{(1)} = \eta^{ab}Q_{ab}^{(2)}
=\eta^{ab}Q_{ab}^{(3)} = 0, \;\;$ and $ e^a \wedge Q_{ab}^{(1)} =0
$. In a similar way the  irreducible decomposition of $ T^a $'s
invariant under the Lorentz group are given in terms of
 \ba
   \alpha =\imath_a T^a \; , \;\;\; \sigma = e_a \wedge T^a
 \ea
so that
 \ba
    T^a = {T^a}^{(1)} +{T^a}^{(2)} +{T^a}^{(3)}      \label{121}
 \ea
where
 \ba
    {T^a}^{(2)} &=& \frac{1}{3}e^a \wedge \alpha \; , \\
    {T^a}^{(3)} &=& \frac{1}{3} \imath^a \sigma \; , \\
    t^a :={T^a}^{(1)} &=& T^a-{T^a}^{(2)} -{T^a}^{(3)} \; .
 \ea
Here $ \imath_a t^a = \imath_a {T^a}^{(3)} = 0, \;\;\; e_a \wedge
t^a = e_a \wedge {T^a}^{(2)} = 0 $. To give the contortion
components in terms of the irreducible components of torsion, we
firstly  write
 \ba
     2K_{ab} = \imath_a T_b - \imath_b T_a
     - (\imath_a \imath_b T_c)e^c             \label{122}
 \ea
from (\ref{contor}) and then substituting (\ref{121}) into above
we find
 \ba
   2K_{ab} &=& \imath_a t_b - \imath_b t_a
               - (\imath_a \imath_b t_c)e^c \nonumber \\
           &{}& +\frac{2}{3} ( e_a \wedge \imath_b \alpha
                - e_b \wedge \imath_a \alpha )   \nonumber \\
           &{}& + \frac{2}{3} ( \imath_a \imath_b \sigma )
 - \frac{1}{3} ( \imath_a \imath_b \imath_c \sigma )e^c.  \label{123}
 \ea
In components $ K_{ab} = K_{c,ab}e^c \; , \;\;\;  t_a =
\frac{1}{2}t_{bc,a}e^{bc} \; , \;\;\; \alpha = F_a e^a \; , \;\;\;
\sigma = \frac{1}{3!} \sigma_{abc} e^{abc} $ this becomes
 \ba
    K_{c,ab} &=& \frac{1}{2} ( t_{ac,b}
                 - t_{bc,a} + t_{ab,c} ) \nonumber \\
           &{}& +\frac{1}{3}( F_b \eta_{ac}
         - F_a \eta_{bc})  -\frac{1}{6} \sigma_{abc}. \label{D.15}
 \ea

\section{Hamiltonian of a Dirac particle in arbitrary space-times}\label{gen-dir}

The Dirac equation in a non-Riemannian space-time with torsion and
non-metricity is written as \cite{kos},\cite{vandyck},\cite{muz2}
 \ba
   ^{*}\gamma \wedge D \psi + M^{*}1 \psi = 0 \label{direqn}
 \ea
in terms of the Clifford algebra ${\mathcal{C}}\ell_{3,1}$-valued
1-forms $ \gamma = \gamma^a e_a $ and  $M =\frac{mc}{\hbar}$. We
use the following Dirac matrices
 \ba
   \gamma^0  &=&  i \left (
                 \begin{array}{cc}
                                   I  & 0  \\
                                   0  & -I
                 \end{array}
          \right )             \; , \;\;\;
   \gamma^i  =  i \left (
                 \begin{array}{cc}
                                   0    &  \sigma^i  \\
                                  -\sigma^i &  0
                 \end{array}
          \right ) \nonumber   \;  \label{dirmat}
 \ea
where $\sigma^i $ are the Pauli matrices. $\psi$ is a  4-component
complex valued Dirac spinor whose covariant exterior derivative is
given explicitly by
 \ba
  D\psi = d\psi + \frac{1}{2}\Lambda^{[ab]}\sigma_{ab} \psi
             + \frac{1}{4}Q \psi \label{covder}
 \ea
where
\ba
     \sigma_{ab} = \frac{1}{4} [ \gamma_a, \gamma_b]
 \ea
are the spin generators of the Lorentz group. We write it out
explicitly as
 \ba
     D\psi &=& \partial_t \psi dx^{\hat{0}} + \partial_\mu \psi dx^\mu +
     \frac{1}{2}e^c \Omega_{c,ab}\sigma^{ab} \psi + \frac{1}{4}
     e^c Q_c \psi \nonumber \\
  &=& {h^{\hat{0}}}_c e^c \partial_t \psi + {h^\mu}_c e^c \partial_\mu
  \psi + \frac{1}{2}e^c \Omega_{c,ab}\sigma^{ab} \psi + \frac{1}{4}
     e^c Q_c \psi
 \ea
where $\Lambda_{[ab]} := \Omega_{ab} = \Omega_{c,ab}e^c$
anti-symmetric part of the full connection 1-form and $Q = Q_a e^a
$ and  using ${ }^*\gamma = \gamma_a { }^*e^a$ and the identity ${
}^*e^a \wedge e^b = - \eta^{ab} { }^*1 $ calculate
 \ba
 { }^*\gamma \wedge D\psi = \left( -{h^{\hat{0}}}_c \gamma^c \partial_t \psi - {h^\mu}_c \gamma^c \partial_\mu
  \psi - \frac{1}{2} \Omega_{c,ab}\gamma^c\sigma^{ab} \psi - \frac{1}{4}Q_c \gamma^c \psi
  \right) { }^*1 \; .
 \ea
Putting this into (\ref{direqn}) we obtain
 \ba
 {h^{\hat{0}}}_c \gamma^c \partial_t \psi = - {h^\mu}_c \gamma^c
 \partial_\mu \psi + M\psi
  - \frac{1}{2} \Omega_{c,ab}\gamma^c\sigma^{ab} \psi - \frac{1}{4}Q_c \gamma^c
  \psi .
 \ea
We multiply this from left by
 \ba
 i\hbar \left( {h^{\hat{0}}}_a \gamma^a \right)^{-1}
 = \frac{-i\hbar}{b^2} \left( {h^{\hat{0}}}_a \gamma^a \right)
 \ea
where
 \ba
 b^2 := \left( {h^{\hat{0}}}_0 \right)^2 +{h^{\hat{0}}}_i
 h^{\hat{0}i} \; .
 \ea
When we compare the result with the Schr\"{o}dinger equation
 \ba
     i\hbar \frac{\partial \psi}{\partial t} = H \psi ,
 \ea
we deduce the Dirac Hamiltonian
matrix~\cite{kon,muz1,muz2,ryder2,ryder,obukhov,hehl2,lammerzahl}
 \ba
    H &=& \frac{c}{b^2} {h^{\hat{0}}}_a {h^\mu}_b \gamma^a \gamma^b  i\hbar \partial_\mu
          -\frac{imc^2}{b^2} {h^{\hat{0}}}_a \gamma^a \nonumber \\
      & & + \frac{i\hbar c}{2b^2} {h^{\hat{0}}}_d \Omega_{c,ab}\gamma^d \gamma^c \sigma^{ab}
           + \frac{i\hbar c}{4b^2} {h^{\hat{0}}}_a Q_b \gamma^a
           \gamma^b  . \label{hamilton}
 \ea
 The right hand side of (\ref{hamilton}) need not be a hermitian matrix
in general; e.g. if ${h^{\hat{0}}}_i \neq 0$, then the mass term
contains an anti-hermitian part such as
 \ba
     H= H_0 +iH_1 \label{H0}
 \ea
where $H_0^+ = H_0$ and $H_1^+ =H_1$.
However, the decomposition (\ref{H0}) is frame dependent. That is
we can always find a local Lorentz frame in which Hamiltonian is
fully hermitian~\cite{ryder,obukhov}. First we can get rid of the
anti-hermitian part of the mass term by diagonalizing the matrix
${h^\alpha}_a$ via a frame transformation
 \ba
     \partial_\alpha (x) \quad & \rightarrow & \quad \partial_\beta (x) {{L^{-1}}^\beta}_\alpha (x) \nonumber \\
   g_{\alpha \beta}(x) \quad  & \rightarrow & \quad  g_{\gamma \delta}(x) {{L^{-1}}^\gamma}_\alpha
   {{L^{-1}}^\delta}_\beta .
 \ea
Thus\footnote{ $L \in SO_+ (1,3)$ where $SO_+ (1,3)$ is special
orthochronous Lorentz group.}
 \ba
 {h^\alpha}_a (x) \quad \rightarrow \quad f(x) \delta^\alpha_a
 \ea
where $x$ stands for $x^\alpha$ and $f(x)$ is composed of
${h^\alpha}_a (x)$. Under this change (\ref{hamilton}) goes over
to
 \ba
 H  & \rightarrow &  H =  cf_1(x) \gamma^0 \gamma^i  i\hbar
                   \partial_{\hat{i}} -imc^2 f_2(x) \gamma^0 \nonumber \\
      & & + i\hbar c f_3(x) \Omega_{c,ab}\gamma^0 \gamma^c \sigma^{ab}
           + i\hbar c f_4(x) Q_b \gamma^0 \gamma^b
 \ea
where $f_i(x)$ are composed of ${h^\alpha}_a (x)$. Putting in the
definition
 \ba
    \Omega_{c,ab} = \epsilon_{abcd}S^d
\ea
and using the identity
\ba
    \gamma^a \sigma^{bc} &=& \frac{1}{2} \eta^{ab}\gamma^c - \frac{1}{2}
    \eta^{ac}\gamma^b - \frac{1}{2}\epsilon^{abcd} \gamma_d
    \gamma_5
 \ea
where $\gamma_5=\gamma_0\gamma_1\gamma_2\gamma_3$, the Hamiltonian
matrix becomes
 \ba
 H  = cf_1(x) \gamma^0 \gamma^i  i\hbar
                        \partial_{\hat{i}} -imc^2 f_2(x) \gamma^0
          + i\hbar c N_a(x) \gamma^0 \gamma^a
                + i\hbar c f_5(x) S_a \gamma^0 \gamma^a \gamma_5
                \label{hamil}
 \ea
where we introduced
 \ba
     N_a := f_3(x) {\Omega^{b,}}_{ba} + f_4(x) Q_a . \nonumber
 \ea
If we now define the canonical momenta
 \ba
      p_i := -i\hbar \left( \partial_{\hat{i}} + \frac{N_i(x)}{f_1(x)} \right)
 \ea
and assume
 \ba
      p_i^+ = p_i ,
 \ea
(\ref{hamil}) takes the form
 \ba
 H  =  f_1(x)cp^i \gamma_0 \gamma_i + imc^2 f_2(x) \gamma_0
                + i\hbar c f_5(x) S^a \gamma_0 \gamma_a \gamma_5
                - i\hbar c N_0(x) \label{51} \; .
 \ea
In order eliminate the last term in (\ref{51}) one may further
perform a locally unitary transformation
 \ba
     \psi (x)\quad  \rightarrow  \quad U^+(x) \psi (x) \quad , \quad
       H  \quad \rightarrow  \quad U^+(x) H U(x)
 \ea
and obtain
 \ba
 H & \rightarrow & H =  f_1(x)cp^i U^+(x)\gamma_0 \gamma_i U(x)
                      + imc^2 f_2(x) U^+(x) \gamma_0 U(x) \nonumber \\
             & &   + i\hbar c f_5(x) S^a U^+(x) \gamma_0 \gamma_a \gamma_5 U(x) \nonumber \\
             & &- i\hbar c [ f_1(x)U^+(x)\gamma_0 \gamma_i \partial^{\hat{i}}U(x) + N_0(x)] \; .
 \ea
Under the following solvable matrix equation
 \ba
    U^+(x)\gamma_0 \gamma_i \partial^{\hat{i}}U(x) =-\frac{
    N_0(x)}{f_1(x)} ,
 \ea
we give the final form of our hermitian Hamiltonian matrix (up to
a sign) by the expression
 \ba
    H  = f_1(x)cp^i \gamma_0 \gamma_i
                      + imc^2 f_2(x)  \gamma_0  + i\hbar c f_5(x) S^a \gamma_0 \gamma_a
                      \gamma_5 \; .
 \ea

\section{Neutrino oscillations in the Kerr background}

Here we construct the Hamiltonian matrix of a Dirac particle (i.e.
a massive neutrino) of mass $m$  in the background space-time
geometry of a heavy, slowly rotating body of mass $M$ such as the
Sun. Its exterior gravitational field will be described by weak
constant, uniform torsion and non-metricity fields, together with
the Kerr metric \cite{kerr}:
 \begin{eqnarray}
 ds^2 = -\left( 1- \frac{2MGr}{c^2 \rho^2} \right) c\, dt \otimes c\, dt + \frac{\rho^2}{\Delta} dr \otimes dr
   +\rho^2 d\theta \otimes d\theta \nonumber \\
   + \left( r^2 +\frac{a^2}{c^2}+\frac{2MGa^2r}{c^4\rho^2} \sin^2\theta
   \right) \sin^2\theta \, d\varphi \otimes d\varphi
   -\frac{4MGar}{c^2\rho^2} \sin^2\theta \, dt \otimes d\varphi
  \end{eqnarray}
where $\Delta = r^2 -\frac{2MG}{c^2}r + (\frac{a}{c})^2$ , $\rho^2
= r^2 + (\frac{a}{c})^2 \cos^2\theta $ , $ a \equiv \frac{J}{M}=
\frac{2}{5} R^2 \omega $. The Sun is assumed a uniform sphere of
radius $R$. $M$, $J$ and $\omega$ are the mass, angular momentum
and angular velocity of the Sun, respectively. We choose the
orthonormal co-frame
 \ba
  e^0 &=& \frac{\sqrt{\Delta}}{\rho}(cdt - \frac{a}{c} \sin^{2}\theta
d\varphi) \quad , \quad  e^1 = \frac{\rho}{\sqrt{\Delta}} dr \nonumber \\
 e^2 &=&  \rho d\theta  \quad \quad \quad \quad , \quad \quad \quad \quad  e^3 = \frac{\sin
\theta}{\rho} \left( \left( r^2 + \left( \frac{a}{c} \right)^2
\right) d\varphi - a dt \right)
 \ea
and  using the definitions
 \ba
     de^a + {\omega^a}_b \wedge e^b =0 \quad \Leftrightarrow \quad
     \omega_{ab} = -\frac{1}{2} \imath_a de_b +
     \frac{1}{2}\imath_b de_a + \frac{1}{2}(\imath_a \imath_b
     de_c)e^c  ,
 \ea
calculate the Levi-Civita connection 1-forms
\begin{eqnarray}
\omega^{0}_{\,\,\,1} &=& \frac{MG[r^2 -
(\frac{a}{c})^2\cos^{2}\theta ] }{\rho^4 c}dt
 + \frac{[(\frac{MG}{c^2}-r)\rho^2 - \frac{2MGr^2}{c^2}]a\sin^{2}\theta}{\rho^4c}  d\varphi ,
\nonumber  \\
\omega^{2}_{\,\,\,3} &=& \frac{2MGra \cos \theta}{\rho^4c^2} dt +
\frac{\Delta (\frac{a}{c})^2 \sin^{2}\theta -( r^2 +
(\frac{a}{c})^2)^2 }{\rho^4} \cos \theta
d\varphi , \nonumber \\
\omega^{0}_{\,\,\,2} &=& - \frac{\sqrt{\Delta}a \sin \theta \cos
\theta}{\rho^2 c} d \varphi ,  \nonumber \\
\omega^{1}_{\,\,\,3} &=& - \frac{\sqrt{\Delta}r \sin
\theta}{\rho^2} d \varphi  , \nonumber \\
\omega^{0}_{\,\,\,3} &=& \frac{\sqrt{\Delta}a \cos
\theta}{\rho^2c} d\theta
-\frac{ar \sin \theta}{\sqrt{\Delta}\rho^2c} dr  , \nonumber \\
\omega^{1}_{\,\,\,2} &=&-\frac{a^2 \sin \theta \cos \theta}{\rho^2
\sqrt{\Delta}c^2} dr - \frac{r\sqrt{\Delta}}{\rho^2} d\theta  .
\label{omega}
\end{eqnarray}

To simplify the discussions,  we  consider only the motion of
massive neutrinos restricted to the equatorial plane of the Sun.
Thus we set $\theta = \pi / 2$ and $d\theta = 0$. Furthermore,
since the Sun rotates very slowly [$\omega \simeq 3 \times 10^{-6}
\; (rad/s)$] we  approximate the metric functions.  Therefore, in
reasonably far away distances from the Sun, the restricted line
element will be taken as
 \begin{eqnarray}
 ds^2 \simeq -\left( 1- \frac{2MG}{c^2 r} \right) c\, dt \otimes c\, dt
 + \left( 1- \frac{2MG}{c^2 r} \right)^{-1} dr \otimes dr \nonumber \\
   + r \, d\varphi \otimes r \, d\varphi
   -4\frac{a}{c} \frac{MG}{c^2r^2} \, cdt \otimes r\, d\varphi \; .
  \end{eqnarray}
We also write the orthonormal co-frame approximately up to $O \, (
\frac{a}{rc})$ as
 \ba
    e^0 = f cdt - \frac{af}{c} d\varphi  \quad , \quad  e^1 = \frac{1}{f} dr
    \quad , \quad e^2 =0 \quad , \quad
    e^3 =-\frac{a}{r} dt + r d\varphi
 \ea
where
 \ba
 \frac{\Delta}{\rho^2} \equiv f^2 \simeq 1-\frac{2MG}{c^2r} \; .
 \ea
The inverses of these relations to the same order of approximation
are
 \ba
  c\,dt = \frac{1}{f}e^0 + \frac{a}{rc} e^3 \quad , \quad  dr = f e^1 \quad , \quad
  d\theta = 0 \quad , \quad  d\varphi = \frac{a}{fr^2c} e^0 +\frac{1}{r} e^3
 \ea
which give
 \ba
     {h^{\hat{0}}}_0 = \frac{1}{f} , \quad {h^{\hat{0}}}_3 = \frac{a}{cr} , \quad
     {h^{\hat{1}}}_1 = f , \quad {h^{\hat{3}}}_0 = \frac{a}{fcr^2} ,
     \quad {h^{\hat{3}}}_3 = \frac{1}{r}
 \ea
 with all the other components neglected.  To this order of
 approximation (\ref{omega}) gives
 \ba
  \omega_{01} \simeq f' e^0 + \frac{a}{cr^2} e^3 , \quad \omega_{03} \simeq  \frac{a}{cr^2}
  e^1,
     \quad \omega_{31} \simeq \frac{a}{cr^2} e^0 + \frac{f}{r} e^3
      \label{omega2}
 \ea
 with the remaining ones neglected.
Then the Hamiltonian matrix (\ref{hamilton}) reads
  \ba
  H &\simeq & f^2 cp_r \gamma_0 \gamma_1 + fc p_{\varphi} \gamma_0 \gamma_3  +ifmc^2 \gamma_0
        -\frac{i}{2}\hbar c ff' \gamma_0 \gamma_1 \nonumber \\
     & &  +\frac{3}{2}i\hbar c f S^a \gamma_0 \gamma_5\gamma_a - i\hbar c f N^a \gamma_0
          \gamma_a \nonumber \\
     & & - \frac{af^3}{r}p_r \gamma_3\gamma_1
         +\frac{ia \hbar f^2f'}{2r} \gamma_3 \gamma_1 - \frac{iamcf^2}{r} \gamma_3 \nonumber \\
      & & +\frac{ia \hbar f}{2r^2}\gamma_0 \gamma_2 \gamma_5
          +\frac{3ia \hbar f^2}{2r} S^a \gamma_3 \gamma_a\gamma_5
         +\frac{ia \hbar f^2}{r}N^a \gamma_3 \gamma_a  \;
 \ea
where
 \ba
 p_r &:=& -i\hbar (\frac{\partial}{\partial r} +\frac{1}{r})\\
 p_\theta &:=& 0 \\
 p_\varphi &:=& -\frac{i\hbar}{r} \frac{\partial}{\partial \varphi} \\
 N_a &:=& \frac{1}{2}({t^b}_{a,b} +F_a + \frac{5}{4}Q_a
    -\Lambda_a)\; . \label{Na}
 \ea
Note that the contributions of axial components of torsion are
given by $S^a$ while certain  components of non-metricity  and the
non-axial components of torsion occur only in $N_a$ and the
rotation effects are given in terms of the parameter $a$. We
rewrite the Hamiltonian $4 \times 4$ matrix in terms of $2 \times
2$ matrices as follows:
 \ba
      H =  \left (
                 \begin{array}{cc}
                                   H_{11} & H_{12} \\
                                   H_{21} & H_{22}
                 \end{array}
          \right )
 \ea
with
 \ba
  H_{11} &=& fmc^2 +iA + B\sigma_1 + (C-i\frac{af^3}{r}p_r)\sigma_2 +D\sigma_3 \nonumber \\
  H_{22} &=&-fmc^2 +iA + B\sigma_1 + (C-i\frac{af^3}{r}p_r)\sigma_2 +D\sigma_3 \nonumber \\
  H_{12} &=& F + (f^2cp_r +iG)\sigma_1 + iH\sigma_2
             +(fcp_\varphi +\frac{amcf^2}{r}+iK)\sigma_3 \nonumber \\
  H_{21} &=& F + (f^2cp_r +iG)\sigma_1 + iH\sigma_2
             +(fcp_\varphi -\frac{amcf^2}{r}+iK)\sigma_3
 \ea
where we set
 \ba
   A & \simeq & -\hbar cf N_0 + \frac{a\hbar f^2}{r}N_3 \nonumber \\
   B & \simeq & \frac{3}{2}\hbar c S_1 + \frac{a\hbar f^2}{r}N_2 \nonumber \\
   C & \simeq & \frac{3}{2}\hbar c S_2 - \frac{a\hbar f^2}{2r^2}(1+rf' +rN_1 ) \nonumber \\
   D & \simeq & \frac{3}{2}\hbar c S_3 - \frac{3a\hbar f^2}{2r}S_0 \nonumber \\
   F & \simeq & \frac{3}{2}\hbar c S_0 - \frac{3a\hbar f^2}{2r}S_3 \nonumber \\
   G & \simeq & -\frac{\hbar cff'}{2r}- \hbar cfN_1 + \frac{3a\hbar f^2}{2r}S_2 \nonumber \\
   H & \simeq & -\hbar cf N_2 - \frac{3a\hbar f^2}{2r} S_1 \nonumber \\
   K & \simeq & -\hbar cf N_3 + \frac{a\hbar f^2}{r}N_0 \;.
 \ea

\bigskip

The way we approach the solar neutrino problem starts by writing
down the Dirac equation in a rotating, axially symmetric
background space-time geometry and  finding phases corresponding
to neutrino mass eigenstates, then finally calculating the phase
differences among them. There are two cases of special interest:
the azimuthal motion and the radial motion. The analysis of the
azimuthal motion with
 $ \vec{p} =(p_r, p_\theta , p_\varphi) = (0,0,p) $ yields for
ultrarelativistic neutrinos, for which $pc \simeq E$ and $cdt
\simeq Rd\varphi$, the phase for the spin up state
 \ba
   \Phi^{\uparrow} = \left( fE + \frac{f m^2 c^4 }{2E}
        + \sqrt{ \Delta_\varphi} +i(A+K)  \right) \frac{R\Delta\varphi}{\hbar c}  \label{spinup}
 \ea
and similarly for the phase of the spin down state
\ba
   \Phi^{\downarrow} = \left( fE + \frac{f m^2 c^4 }{2E}
        - \sqrt{ \Delta_\varphi} +i(A+K)  \right) \frac{R\Delta\varphi}{\hbar c} \label{spindown}
 \ea
 where
 \ba
 \Delta_\varphi \simeq B^2 +C^2 +D^2 +F^2+G^2+H^2+ 2 (DF +BH -CG) \; . \label{3.27}
   \ea
These phases alone do not have an absolute meaning; the quantities
relevant for the interference pattern at the observation point of
the neutrinos are the phase differences $ \Delta \Phi = \Phi_2 -
\Phi_1 $ where $\Phi_1$ and $\Phi_2 $ are the absolute phases of
the neutrino mass eigenstates $ \nu_1 $ and $ \nu_2 $. It is thus
seen from equations (\ref{spinup}) and (\ref{spindown}) that the
phase differences can have explicit dependence on non-metricity in
the case of opposite spin polarizations of mass eigenstates for
the azimuthal motion via (\ref{3.27}):
 \ba
  \Delta \Phi = \Phi^{\downarrow}_2 - \Phi^{\uparrow}_1 = \left( \frac{\Delta m^2 c^4}{2(E/f)}
                -2\sqrt{ \Delta_\varphi } \right) \frac{R \Delta \varphi}{\hbar c} , \\
  \Delta \Phi = \Phi^{\uparrow}_2 - \Phi^{\downarrow}_1 = \left( \frac{\Delta m^2 c^4}{2(E/f)}
                +2\sqrt{ \Delta_\varphi } \right) \frac{R \Delta \varphi}{\hbar c}
 \ea
where $\Delta m^2 = {m_2}^2-{m_1}^2$.

The Hamiltonian for the radial motion on the other hand is
obtained by the assumption $\vec{p} = (p,0,0) $. In this case with
the further assumptions $ pc \simeq E$  and $cdt \simeq dr$, the
phases appropriate to the spin up and spin down particles are,
respectively,
 \ba
 \Phi^{\uparrow} &=& \frac{1}{\hbar c} \int{\left( f^2 E + \frac{m^2c^4}{2E}
                    + \sqrt{\Delta_r} +i(A+G) \right) }dr , \label{eq24} \\
  \Phi^{\downarrow} &=& \frac{1}{\hbar c} \int{\left( f^2 E + \frac{m^2c^4}{2E}
                    - \sqrt{\Delta_r} +i(A+G) \right) }dr \label{eq25}
 \ea
where
 \ba
 \Delta_r &\simeq& (D-H)^2 +(B+F+\frac{amH}{rp})^2 +(C+K-\frac{amG}{rp})^2 \nonumber \\
          & &-\frac{a^2f^4}{r^2}(mc-fp)^2 +\frac{2iaf^2}{r}(mc-fp)(C+K-\frac{amG}{rp})  \;
          . \label{Delta-r}
 \ea
In this case the relevant phase differences depending on
non-metricity via $N^a$ and rotation via $a$ come from the
opposite spin polarization states
 \ba
  \Delta \Phi &=& \Phi^{\downarrow}_2 - \Phi^{\uparrow}_1 =
      \frac{\Delta m^2 c^3}{2\hbar E} \Delta r
    - \frac{2}{\hbar c} \int{ \sqrt{\Delta_r} }dr ,\label{82}\\
 \Delta \Phi &=& \Phi^{\uparrow}_2 - \Phi^{\downarrow}_1 =
\frac{\Delta m^2 c^3}{2\hbar E} \Delta r
    + \frac{2}{\hbar c} \int{ \sqrt{\Delta_r} }dr \label{83}\; .
 \ea
We point out that $\Delta_r = \mbox{Re}\Delta_r
+i\mbox{Im}\Delta_r$ implies $\sqrt{\Delta_r} = \alpha +i\beta$
and hence the rotation of the Sun would suppress the transitions
among the neutrinos via the phase difference equations
(\ref{82}),(\ref{83}) in opposite spin polarizations.

\bigskip

\section{Conclusion}

 \noindent
We have here extended our recent study of gravitationally induced
neutrino oscillations~\cite{muz1} by including the effects of
rotation of the Sun, space-time non-metricity and  as well as
components of torsion other than the axial ones. The rotation of
the Sun implies a damping of neutrino oscillations. However, this
result is frame dependent as we explained in Sect.\ref{gen-dir} in
general. We have shown that there are contributions coming from
non-axial components of spacetime torsion and definite components
of spacetime non-metricity depending on the polarizations of the
spin states of the mass eigenstates. If we set the rotation
parameter  $a=0$, then (\ref{Delta-r}) gives
 \ba
     \sqrt{\Delta_r} = \frac{3}{2} \hbar c \left( (S_0 +S_1)^2 +(S_2 -\frac{2}{3}fN_3)^2
                +(S_3 +\frac{2}{3}fN_2)^2 \right)^{1/2}
 \ea
which means that there is no suppression among the neutrinos and
only $N_2$ and $N_3$ components of $N^a$ contribute to the
oscillations. If we further  set $N^a=0$, we reach agreement with
our previous results in  \cite{muz1}. It should be clear that the
above scheme only works if the neutrino masses are different from
in each other and hence, in general, different from zero. This
means there are right-handed neutrinos as well as left-handed ones
which, however, must interact with matter very weakly as they have
not yet been observed.
 Finally, we note that all the possible contributions discussed
here so far would be of the order of Planck scales, and hence do
not suffice to account for the observed solar neutrino deficit.

\bigskip
\section{Acknowledgement}

 \noindent We thank the referees for constructive criticisms.
One of the authors (MA) acknowledges partial support through the
research project BAP2002-FEF007 by  Pamukkale University, Denizli.

\end{document}